\documentclass[prb,preprint,showpacs,floatfix]{revtex4}

\usepackage{graphicx}
\usepackage{array}
\usepackage{amsmath,amsfonts,amssymb}
\usepackage[ansinew]{inputenc}
\usepackage{color}
\usepackage{calc}
\usepackage[normalem]{ulem}

\begin{document}

\title{Topologically localized excitons in single graphene nanoribbons}
\author
{Song Jiang$^{1\ast}$, Tom{\'a}\v{s} Neuman$^{1,2}$, Alex Boeglin$^1$, \\
 Fabrice Scheurer$^1$, Guillaume Schull$^{1\ast}$\\
\normalsize{$^1$ Universit\'e de Strasbourg, CNRS, IPCMS, UMR 7504, F-67000 Strasbourg, France,} \\
\normalsize{$^2$ Institut des Sciences Mol\'eculaires d'Orsay (ISMO), UMR 8214, CNRS, Universit\'e Paris-Saclay, 91405 Orsay Cedex, France.}\\
\altaffiliation{guillaume.schull@ipcms.unistra.fr}
\altaffiliation{song.jiang@ipcms.unistra.fr}
}

\begin{abstract}
Excitonic emission from atomically precise graphene nanoribbons (GNRs) synthesised on a metal surface is probed with atomic-scale spatial resolution using a scanning tunneling microscopy (STM) approach. A STM-based strategy to transfer the GNRs to a partially insulating surface is used to prevent light emission quenching of the ribbons by the metal substrate. Sub-nanometer resolved STM-induced fluorescence spectra reveal emission from localized dark excitons build upon the topological end states of the GNRs. A low frequency vibronic emission comb whose characteristics change with the GNR length is attributed to longitudinal acoustic modes confined to a finite box. Overall, our study provides a novel path to investigate the interplay between excitons, vibrons and topology in atomically precise graphene nanostructures.  
\end{abstract}
\date{\today}

\maketitle
\begin{figure}
\includegraphics[width=14.5cm]{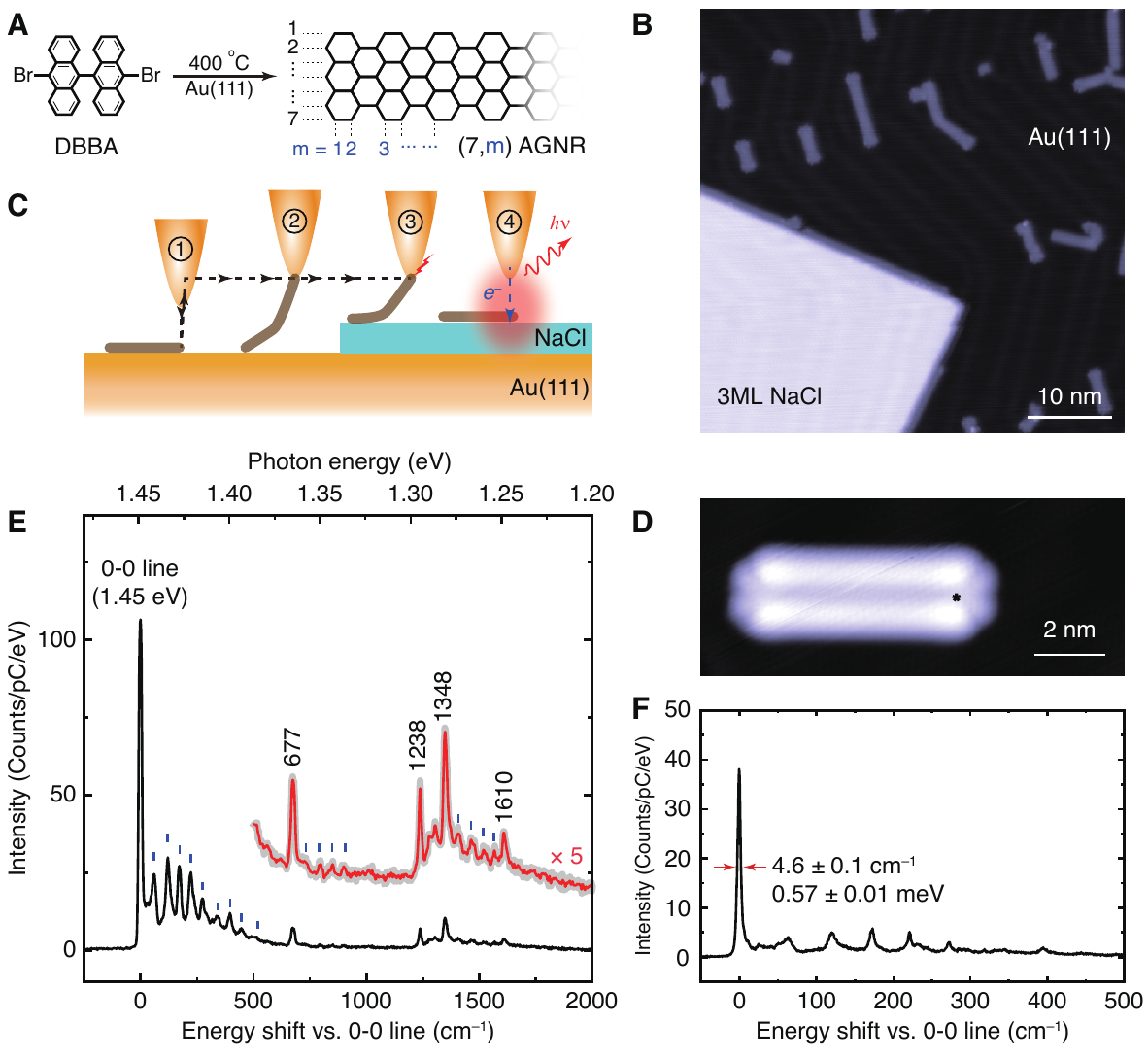}
\caption{\label{fig1} \textbf{STM-induced luminescence from decoupled ($7,m$)AGNR}.
(\textbf{A}) On-surface synthesis of ($7,m$)AGNRs from molecular precursor DBBA. (\textbf{B}) Typical STM image ($V$= 0.05 V, $I$ = 5 pA) of ($7,m$)AGNRs after subsequent deposition of 3ML-NaCl islands on Au(111). (\textbf{C}) Sketch of the STM manipulation procedure to transfer ($7,m$)AGNRs from the Au(111) surface to a 3ML-NaCl island: (1) picking up the ribbon at one terminus with the STM tip;  (2) laterally moving the ribbon to the 3ML-NaCl; (3) releasing the ribbon from the tip with a bias pulse ($V$ = 3.5 V, duration = 30 ms); (4) positioning the tip on the ribbon to measure STML spectra.  (\textbf{D}) STM image ($V$ = $-$2.3 V, $I$ = 5 pA) of a ($7,28$)AGNR transferred onto a 3ML-NaCl island through STM manipulation. (\textbf{E}) STML spectrum ($V$ = 2.3 V, $I$ = 100 pA, t = 120 s) acquired for the tip located on one end (black dot in \textbf{D}) of the decoupled ($7,28$)AGNR.  (\textbf{F})  Highly resolved STML spectrum ($V$ = 2.3 V, $I$ = 200 pA, $t$ = 120 s) acquired from the same decoupled ($7,28$)AGNR with a 1200 grooves/mm grating and a 5 $\mu$m slit (corresponding to a spectral resolution of 0.35 meV at 1.45 eV).} 
\end{figure}
\begin{figure}
\includegraphics[width=14.5cm]{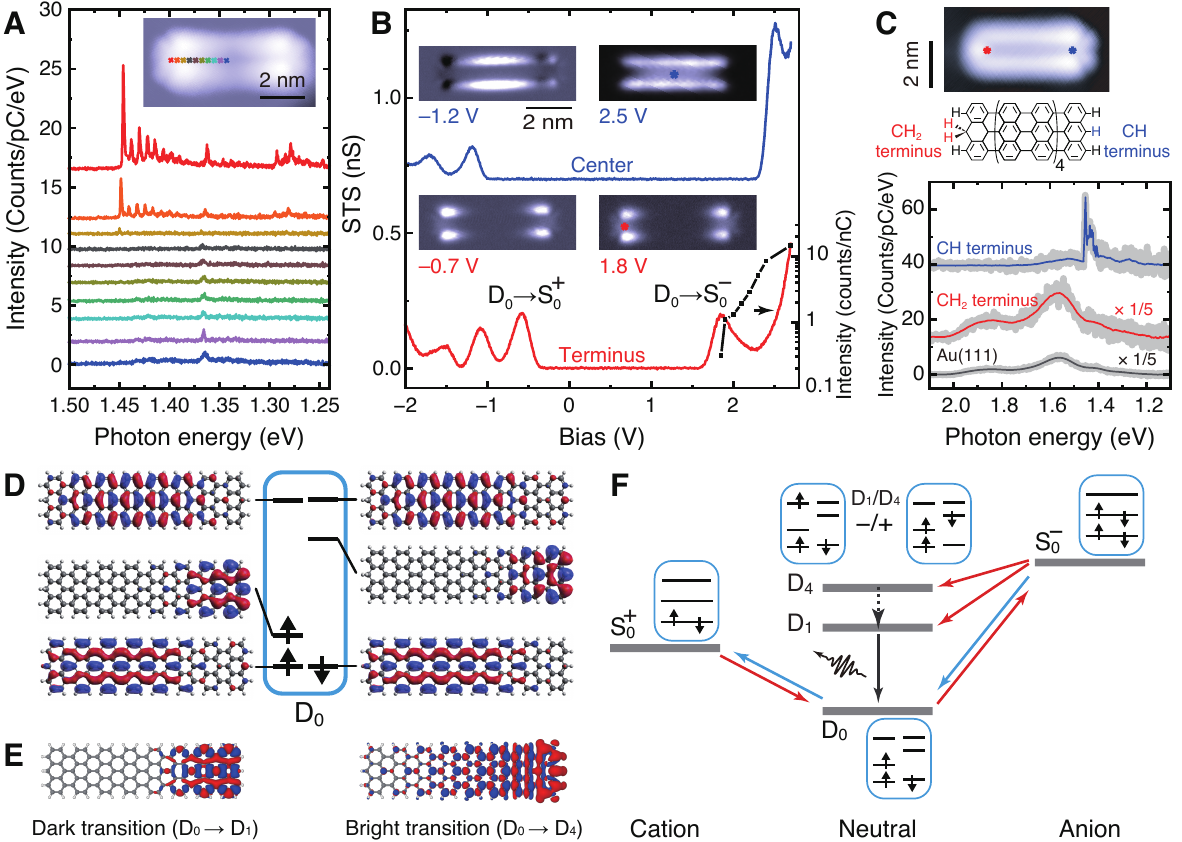} 
\caption{\label{fig2} \textbf{Local excitonic emission from ($7, m$)AGNR}.
(\textbf{A}) STML spectra ($V$= 2.7 V, $I$ = 200 pA, $t$= 300 s) acquired over the line running along the long axis of a decoupled ($7, 24$)AGNR imaged in inset ($V$= 2.3 V, $I$= 3 pA). (\textbf{B}) d$I$/d$V$ spectra acquired in the center (blue) and at one extremity (red) of the ($7, 24$)AGNR imaged in \textbf{A}. Constant height d$I$/d$V$ maps acquired at voltages corresponding to d$I$/d$V$ resonances are displayed in insets. The voltage dependency of the 0-0 emission efficiency appears as black dots overlaid on the red spectrum.  (\textbf{C}) STM image ($V$= $-$2.5 V, $I$= 5 pA) and sketch of a decoupled ($7, 20$)AGNR with one CH terminus (\textit{i.e.}, leading to a $sp^2$ carbon atom) and one CH$_2$ terminus (\textit{i.e.}, leading to a $sp^3$ carbon atom), and STML spectra (($V$= 2.0 V) acquired on each terminus and on top of the bare Au(111). 
(\textbf{D}) Frontier Kohn-Sham orbitals and their corresponding ground-state occupations for both spin channels. (\textbf{E}) Transition electron density associated to the D$_{0}\to$D$_{1}$ (left) and D$_{0}\to$D$_{4}$ (right) transitions calculated using TDDFT. (\textbf{F}) Fluorescence excitation model: at sufficiently high positive voltage (1.85 V) the GNR can be transiently driven in its negative charge state (S$_{0}^{-}$) by charge tunneling from the tip to one of its topological end states. Subsequent tunneling of this charge to the substrate may leave the GNR in the excited neutral state (D$_1$ to D$_4$) that non-radiatively relax to the lowest lying state D$_1$. The molecule eventually relaxes in its ground state D$_0$ by emitting a photon. 
}
\end{figure}

Since their first on-surface synthesis,\cite{cai2010atomically} atomically precise graphene nanoribbons (GNRs) have attracted a tremendous interest in the nano-science and technology communities for their topology-related physical properties.\cite{Bo2021Novel,Houtsma2021Atomically,Miao2021The,wang2021graphene,Song2022} Indeed, their specific edge conformations host peculiar electronic states that in turn lead to unconventional transport or magnetic properties. \cite{koch2012voltage, ruffieux2016surface, groning2018engineering, rizzo2018topological, Daniel2020Inducing, blackwell2021spin} Their optical properties, on the other hand, hold great promises towards robust and controllable atomically thin optoelectronic devices.\cite{Liu2022Small} In fact, GNRs combine many of the outstanding characteristics of graphene with an electronic gap, a mandatory property for many applications including light-emitting devices. Whereas theoretical studies discuss in great details how the optical properties of GNRs may be advantageously controlled through atomic-scale variations of their  width, length and edge shapes,\cite{Yang2007Excitonic, Prezzi2008Optical, Prezzi2011Quantum, Alfonsi2012Excitonic, Attaccalite2017Excitonic, Deilmann2017Huge, cardoso2018termini, Wu2020Controlling} experiments reporting on the excitonic properties of GNRs are scarce,\cite{denk2014exciton, soavi2016exciton, Bronner2016Excitonic, Barin2019Surface, Zhao2020Optical, Tries2020Experimental} especially those focusing on fluorescence of on-surface grown GNRs. Here, experiments are limited to ensemble averaging measurements where defects are shown to play a significant  role,\cite{Senkovskiy2017Making, Pfeiffer2018Observation, Ma2020Engineering} or to individual elements in direct contact with metallic electrodes altering the GNR excitonic properties.\cite{chong2018bright} Indeed, because the synthesis of these GNRs is performed directly on metallic surfaces -- which in turn causes luminescence quenching -- the intrinsic emission properties of atomically precise GNRs remain so far an almost unexplored territory. \\
In this manuscript, we build on a strategy that consists in using a STM-tip to transfer individual GNRs -- being 7 atom-wide and having armchair edge structures (7-AGNRs) -- from the bare part of a Au(111) surface to a neighbouring thin insulating NaCl layer.\cite{wang2016giant} Using STM-induced luminescence (STML) we could then address the fluorescence properties of single GNRs isolated from any contact with metallic electrodes. Our STML data reveal a sharp (FWHM $\leq$ 0.6 meV) emission line at an energy that is lower than the excitonic emission expected for an infinitely long ribbon, and that is traced back to dark excitons localized at the topological state of the GNR termini. This emission line is accompanied by an extremely rich and complex vibronic emission spectrum. Among the vibronic features, a series of intense equally spaced peaks, whose energy separation scales with the inverse of the GNR length, stands out. It is associated to the coupling between the localized excitons with longitudinal acoustic modes confined within the finite-length GNR. These features eventually merge and form a band for longer ribbons, revealing the transition of the exciton-phonon coupling from the discrete to the continuous limit.  Our study suggests a path based on chemical engineering of topological states to control the emission properties of GNRs. Readily applicable to any other types of GNRs, our approach opens a new era in the study of the optical properties of this promising material.\\
Recent STML works have demonstrated that one can excite the fluorescence properties of individual organic molecules when they are sufficiently decoupled from a metallic substrate.\cite{Qiu2003Vibrationally, zhang2016visualizing, imada2016real, doppagne2018electrofluorochromism, dolezal19, kimura2019selective} A common strategy consists in evaporating the molecule on a thin insulating layer of oxide or salt adsorbed on a metal surface. This efficient approach is however not suited to long organic structures such as graphene nanoribons whose synthesis requires a catalytic reaction step at the surface of metallic substrates.\cite{Clair2019Controlling} In Fig. \ref{fig1} we detail our method to address the luminescence properties of single GNRs with sub-nanometer scale precision. We first follow the usual on-surface synthesis approach (Fig. \ref{fig1}A) to form $7$ atom-wide and $m$  atom-long armchair graphene nanoribons (($7, m$)AGNR) on a Au(111) surface from a 10,10'-dibromo-9,9'-bianthryl (DBBA) precursor,\cite{cai2010atomically} and subsequently evaporate NaCl so as to form 3 monolayer-thick (3ML) NaCl islands on Au(111) (see Supplementary Information(SI) Section I for more details). A STM image of the substrate after such a preparation is provided in Fig. \ref{fig1}B where one can identify a clean NaCl island (bottom left) and several ($7,m$)AGNRs of different lengths and orientations dispatched on the bare gold area. In Fig. \ref{fig1}C we schematically explain how the tip of the STM is used to transfer a ($7,m$)AGNR adsorbed on the gold surface onto a NaCl cluster\cite{wang2016giant}  (see SI section II for details): (1) the tip is approached to a ($7,m$)AGNR extremity until contact is reached. A weak bond between the last tip atom and the reactive ribbon terminus allows us to lift the ribbon in the junction by retracting the tip by a few nanometers; (2) the tip is then laterally displaced on top of a NaCl cluster where (3) the ribbon is released by applying a voltage pulse of 3.5 V amplitude and 30 ms width. In Fig. \ref{fig1}D a STM image of a ($7,28$)AGNR deposited on 3ML NaCl following this method is provided. No modification of the ribbon can be observed, indicating that the procedure does not affect the structure of the GNR. \\
In Fig. \ref{fig1}E we display a STML spectrum acquired for the STM tip located at the position marked by a black dot in Fig. \ref{fig1}D. This spectrum reveals an intense and complex signal composed of sharp lines of excitonic nature, attesting for the success of our decoupling procedure. In this STML spectrum, one first identifies an intense 0-0 line at $\approx$ 1.45 eV (855 nm), with a sub-meV spectral width (Fig. \ref{fig1}F), followed by several features of weaker intensities assigned to vibronic emission. Whereas the vibronic peaks at high energy ($>$~1000 cm$^{-1}$) are reminiscent of Raman-like patterns frequently observed in STML spectra,\cite{Dop07Vibronic,doppagne2018electrofluorochromism,kong2021probing} the series of equally spaced peaks at low energy ($<$~500 cm$^{-1}$) is exceptional. It is discussed in the second part of the manuscript. The energy of the 0-0 line (h$\nu$ = 1.45 eV) is intriguingly low, as the lowest excitonic transition is expected at $\approx$ 2 eV for a ($7,\infty$)AGNR.\cite{denk2014exciton} However, no fluorescence  contribution is observed at higher energy, independently of the used voltage bias (up to 2.8 V) or of the GNR length.\\ 
To identify the origin of the 0-0 line, we display in Fig. \ref{fig2}A a series of STML spectra recorded along the main axis of a decoupled ($7,24$)AGNR. At the exception of a peak at 677 cm$^{-1}$ (discussed later), all spectral contributions fade rapidly when the tip is moved away from the ribbon terminus. A differential conductance (d$I$/d$V$) spectrum recorded at a GNR terminus (red spectrum in Fig. \ref{fig2}B) reveals electronic states at $V$ = $-$0.6 V and $V$ = 1.9 V corresponding to localized states of topological nature that are absent from spectra recorded at the center of the ($7,24$)AGNR (blue spectrum in Fig. \ref{fig2}B). These topological end states result from the unsaturated electronic structure of the $sp^2$ hybridized carbon atoms located at the center of the zigzag  ($7, 24$)AGNR termini.\cite{wang2016giant} The similar spatial dependencies of the optical and electronic signals suggest that the topological end states are involved in the fluorescence process. To confirm this hypothesis, we investigate the STML properties of a decoupled ribbon (Fig. \ref{fig2}C) having the central carbon atom of one of its termini bonded with two hydrogen atoms (labeled as ``CH$_2$ terminus''). This configuration, which naturally occurs for a fraction of the ($7, m$)AGNRs synthesized on Au(111) surfaces, is known to saturate the ribbon electronic structure, leading to a $sp^3$ hybridization of the central carbon atom and to the absence of topological-state on this side\cite{Lit13Suppression, talirz2013termini}. The STML spectrum (in red) acquired at this CH$_2$ terminus reveals broad emission resonances similar to the plasmonic emission measured with the same tip on top of the Au(111) substrate (in black). In contrast, the spectrum acquired at the opposite CH terminus (in blue) reveals an excitonic emission signature. All these observations indicate a prominent role of the topological end-states in the fluorescence process. \\
To elucidate the role of these end states we perform time-dependent density-functional theory (TDDFT) calculations of a ($7, 16$)AGNR whose left edge is ``saturated'' as described above. Due to the open-shell nature of the electronic structure of this half-saturated ribbon, the ground-state density is obtained from a spin unrestricted doublet DFT calculation. We plot the corresponding frontier Kohn-Sham orbitals and their ground-state occupations for both spin channels in Fig. \ref{fig2}D. The three orbitals on the left panel are shown for the majority spin [up], and those on the right panel for the minority spin [down]. The Kohn-Sham energies corresponding to these orbitals are schematically shown in the central panel. The pair of occupied orbitals (bottom) is reminiscent of the valence band of infinite GNRs; conversely, the two unoccupied orbitals (top) correspond to the GNR conduction band. The singly-occupied and unoccupied orbitals (middle) are clearly localized on the unsaturated (right) edge (Fig. \ref{fig2}D) and can be associated with the topological states. \\
We next calculate the excited states of the ribbon using linear-response TDDFT in the random-phase approximation as implemented in Gaussian 16\cite{g16} (more details in SI section III). We identify low-lying excited states that can be seen as linear combinations of configurations where a spin-up electron, promoted from the edge state, appears in the conduction band, and a spin-down electron, promoted from the valence band, appears in the empty edge state, in agreement with earlier calculations.\cite{chong2018bright} The transition of lowest energy (1.59 eV) carries a negligible transition dipole moment, resulting from the destructive interference between the spin-up and spin-down transition channels, and we therefore denote this transition as dark. The absence of a transition dipole moment becomes apparent upon inspection of the corresponding transition charge density (D$_0 \rightarrow$ D$_1$) shown in Fig.\,\ref{fig2}E, left, which also demonstrates that the excitation is localized on the ribbon's non-saturated terminus. \\
Interestingly, the oscillator strength of this dark state can be activated through the efficient coupling with the picocavity plasmon confined at the tip, as was suggested in previous works. \cite{zhang2016visualizing, neuman2018nanolett} We therefore assign this transition to the experimentally observed 0-0 peak at 1.45 eV. This assignment is also consistent with the particularly narrow width of the 0-0 line that reflects the long-lived nature of the D$_1$ dark state. On the other hand, the same spin-dependent transition channels can also interfere constructively and give rise to a bright transition between the ground state and a higher-lying excited state D$_4$ (1.82 eV). Due to the involvement of an orbital localized close to the non-saturated end of the ribbon in this excitation, its transition density (D$_0 \rightarrow$ D$_4$) is also localized on this terminus (Fig. \ref{fig2}E, right). Other dark excitations appear in the TDDFT calculations between D$_1$ and D$_4$, and are discussed in SI section III.\\
Building on our TDDFT calculations and experimental observations, in Fig. \ref{fig2}F we propose a model based on a many-body representation of the GNR states to explain the mechanisms of the reported GNR fluorescence. If one considers only what happens on one side of the ribbon, the ($7, m$)AGNR on NaCl/Au(111) is in a neutral ground state of doublet character, D$_{0}$. At a negative voltage of $\approx$ $-$0.6 V, an electron can tunnel from the ($7,m$)AGNR to the tip (blue arrow), driving the ribbon into a positively charged state of singlet character S$_{0}^{+}$. This state is only transiently populated as the ($7, m$)AGNR is rapidly neutralized back to its original state  (D$_{0}$) by tunneling of an electron from the substrate (red arrow). The topological end-state resonance at $V$ = $-$0.6 V in Fig. \ref{fig2}B (red spectrum) therefore corresponds to the D$_{0} \rightarrow$ S$_{0}^{+}$ transition. The same reasoning applies for the other end-state resonance at $V$ = + 1.9 V, which therefore corresponds to a D$_{0} \rightarrow$ S$_{0}^{-}$ transition and a transiently negatively charged ($7, m$)AGNR. Here as well, the ribbon may rapidly return to the initial D$_{0}$ state by tunneling of an electron to the substrate. But as S$_{0}^{-}$ has an absolute energy (1.9 eV) that is higher than the one of the excited states (D$_1$ to D$_4$), S$_{0}^{-} \rightarrow$ D$_{1}$ to D$_{4}$ transitions may occur. The fact that eventually only the D$_{1} \rightarrow$ D$_{0}$ emission is observed in the experiment indicates fast non-radiative transitions from D$_{i}$ ($i\geq 2$) to the lowest excited dark state D$_1$, and explains why the fluorescence of the 7AGNR appears intrinsically low in usual photoluminescence measurements.\cite{Senkovskiy2017Making} The bias onset of the D$_{1} \rightarrow$ D$_{0}$ emission ($\approx$ 1.85 V; black squares in Fig. \ref{fig2}B and section IV in SI) matches the voltage required to tunnel into the S$_{0}^{-}$ state, confirming the proposed mechanism.\\  
\begin{figure}
\includegraphics[width=16cm]{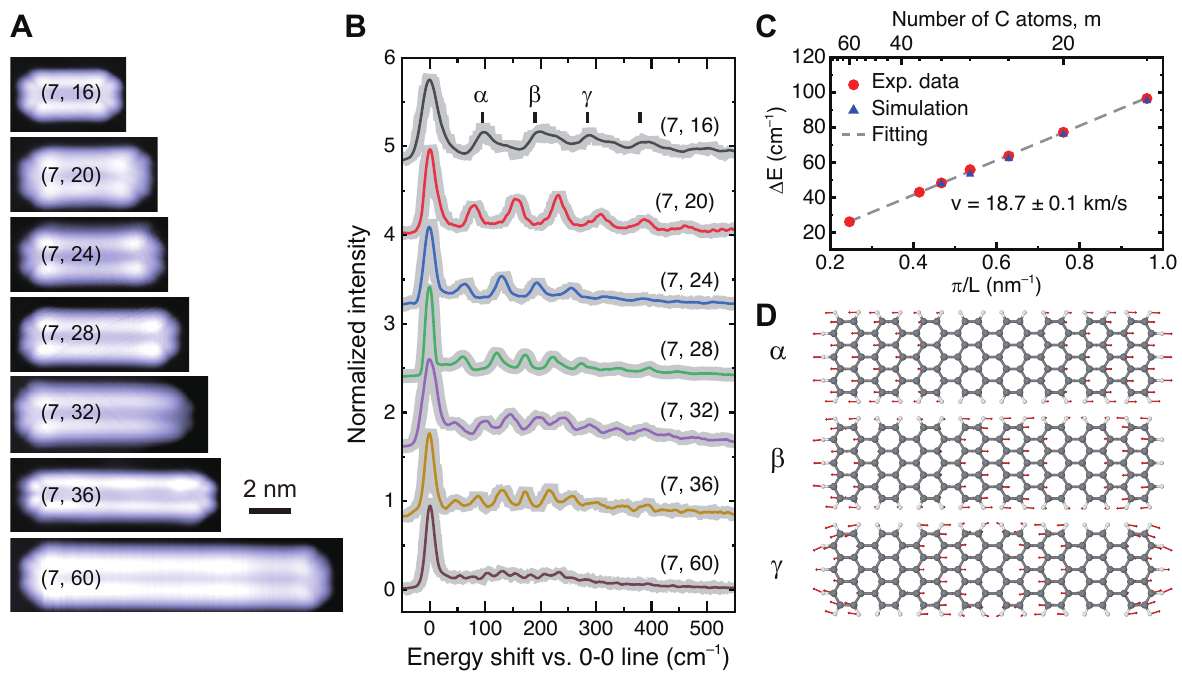} 
\caption{\label{fig3} \textbf{Coupling of excitons with longitudinal acoustic modes for ($7, m$)AGNRs of increasing length}.
(\textbf{A}) STM images ($-$2 V $< V < -$3 V) of decoupled ($7, m$)AGNRs.  (\textbf{B}) STML spectra (2.3 V $< V <$ 2.7 V) acquired from the ribbons in \textbf{A}. (\textbf{C}) experimental (red dots) and calculated (blue triangles) average energy separation, $\Delta$E, between successive vibronic peaks as a function of $m$. (\textbf{D}) Three first LAM modes ($\alpha$, $\beta$, $\gamma$) calculated by DFT for a ($7, 16$)AGNR. The arrows indicate the normalized atomic displacement profile (the arrow lengths were scaled by a factor 10).
 }
\end{figure}
A remarkable advantage of GNRs over usual chromophores is that one can envisage to tune their optoelectronic properties by changing their length. In Fig. \ref{fig3} we investigate how this parameter affects the STML properties of ($7, m$)AGNRs by studying ribbons made out of four (($7, 16$)AGNR) to fifteen (($7, 60$)AGNR) DBBA units (Fig. \ref{fig3}A). For all these ($7, m$)AGNR, the energy of the 0-0 line remains essentially constant (more details in SI section V), in agreement with an optical transition determined by excitons localized at the ribbon termini. The series of vibronic peaks at low energy ($<$ 500 cm$^{-1}$) presents a radically different behaviour. One can see in Fig. \ref{fig3}B that the energy separation between successive peaks decreases with ribbon length, reaching a limit for the ($7, 60$)AGNR where separating the peaks from each other is tedious. Besides, the number of peaks in the comb increases up to $\approx$ 10 for the longest ribbons. These behaviours reflect the confinement of acoustic modes, hereafter referred to as longitudinal acoustic modes (LAM), in the GNRs that act as one-dimensional boxes of controllable length. The first order LAM  mode ($\alpha$) was identified in Raman measurements performed on ensembles of length-selected AGNRs,\cite{overbeck19A} but it is the first time that the higher-order modes are reported and that their dispersion can be followed as a function of the GNR length. In Fig. \ref{fig3}C we report the evolution of the energy separation between successive vibronic lines as a function of the number of carbon atoms $m$ in the long axis of the ribbons. As expected, Fig. \ref{fig3}C reveals a linear dispersion of the modes with a slope corresponding to the speed of sound ($v$ = 18.7 km/s) in GNRs and  graphene.\cite{Gillen2010Raman, overbeck19A, CONG201919Probing} Remarkably, DFT simulations (blue triangles in Fig. \ref{fig3}C) of the different ($7, m$)AGNRs reproduce almost perfectly the dispersion observed in our STML spectra. This confirms the vibronic peak assignment and allows us to represent  the first three LAM modes ($\alpha$, $\beta$ and $\gamma$ in Fig. \ref{fig3}D) for a ($7,16$)AGNR.\\
Interestingly, the envelope of the vibronic comb is very similar for all ribbons, showing attenuated emission for the peaks closest to the 0-0 line, a maximum intensity reached for peaks from 100 to 200 cm$^{-1}$, and a slow decay at higher energy. At the same time, the maximum intensity of the vibronic peaks reaches only up to 50\% of the 0-0 line intensity. In a Franck-Condon (FC) picture, multiple excitations of the same vibrational mode would also result, as in Fig. \ref{fig3}, in regularly spaced vibronic peaks.\cite{Qiu2003Vibrationally} However, the overall shape of the vibronic comb (\textit{i.e.}, a dominant 0-0 line and vibronic peaks whose intensity increases and later decreases with detuning from the 0-0 line) cannot be reconciled with a FC distribution. The data of Fig. \ref{fig3}, therefore indicate that each vibronic peak of the comb corresponds to single excitations of LAMs of different orders. Besides, only modes having even symmetry (such as $\alpha$ and $\gamma$) have been predicted to be optically active in far-field Raman spectroscopy, \cite{overbeck19A} as only those result in a change of polarizability. This selection rule is clearly not respected in our experiments, as all the modes, regardless of their even/odd quality, appear intense in the spectra. This somewhat surprising behaviour can be assigned to the localization of the exciton to only one end of the GNR (\textit{i.e.}, lowering the symmetry of the system) in our STML experiment. This results in a net oscillator strength for modes of both even and odd symmetries. Moreover, fitting the relative intensities of the LAMs with a FC model (described in SI section VI) allows us to retro-engineer the effective deformation associated to the exciton creation. These data suggest that the exciton confinement is even stronger than expected from the gas-phase TDDFT calculations, an effect that is most likely due to the presence of the tip in our experiment. The overall emerging picture is that of emitters strongly localized at the topological ends of the GNRs, and Franck-Condon coupled to delocalized acoustic modes of the ribbon structure. The spectroscopic vibronic pattern reflects here an end-localized deformation in the excited state with respect to the ground-state geometry.
\begin{figure}
\includegraphics[width=16 cm]{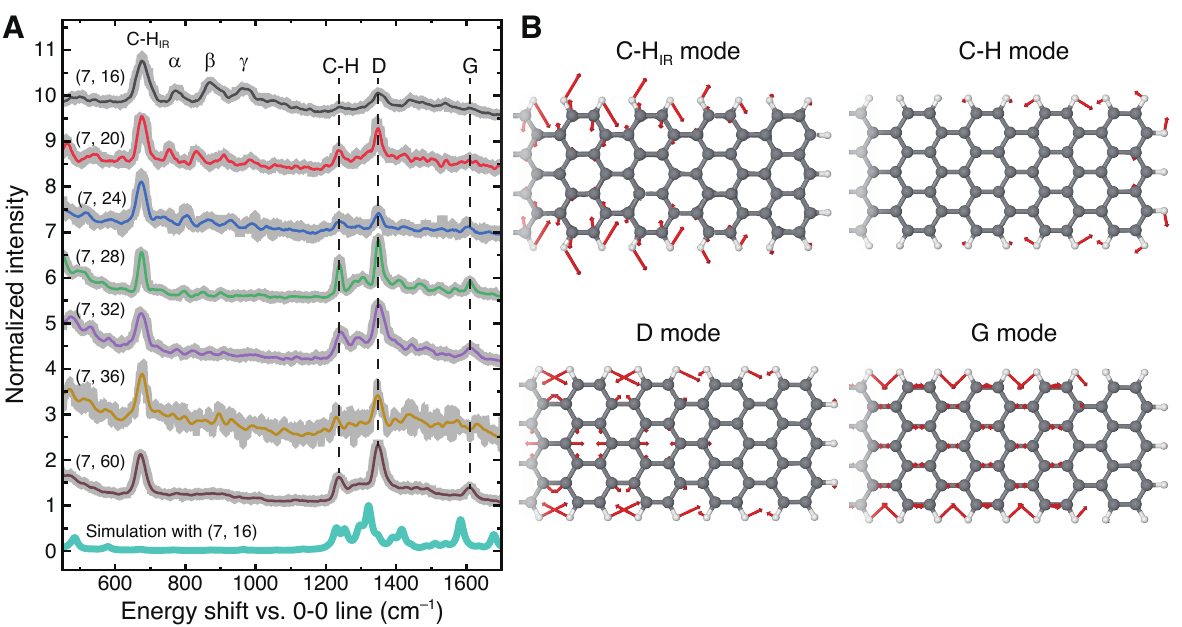}
\caption{\label{fig4} \textbf{Fingerprint region of the ($7, m$)AGNR vibronic spectra}.
(\textbf{A}) High energy region of the STML spectra (2.3 V $< V <$ 2.7 V) acquired from the same ribbons imaged in Fig. \ref{fig3}\textbf{A}. (\textbf{B}) DFT calculation of the main modes identified in the spectra in \textbf{A}. 
A visual representation of the high-frequency vibrational modes calculated for the ($7, 16$) ribbon (only half a ribbon is shown) that can be tentatively assigned to the vibronic peaks obtained in the spectra in \textbf{A}. The vectors visualizing the vibrational modes (normalized atom displacements) have been scaled by a factor 10 for C-H$_{\rm IR}$, D and G modes, and by a factor 5 for the C-H mode.}
\end{figure} 
At last, we discuss the length-dependency of the high ($>$ 500 cm$^{-1}$) energy vibronic peaks (Fig. \ref{fig4}). Extremely characteristic of the probed material, this spectral section is often referred to as the fingerprint region. In contrast with the low energy peaks, the most intense peaks here do not shift with ribbon length. All, at the exception of the 677 cm$^{-1}$ peak, can be identified based on a comparison with calculations of the vibronic activities (Fig. \ref{fig4}A) and literature.\cite{Fedotov2020Excitonic} The 1238 cm$^{-1}$ can be assigned to in-plane bond-bending vibrational modes of the C-H bonds at the GNR edges.  The 1348 cm$^{-1}$ and 1610 cm$^{-1}$ are the so-called D- and G-modes respectively associated to E$_{\rm 2g}$ and A$_{\rm 1g}$ deformations of the carbon rings that one finds in all vibronic spectra of $sp^2$ carbon materials. Their intensity ratio D/G is generally used to determine the presence of defects in meso- and macro-scopic scale systems, and a low number assesses for an overall good structural quality of the material. \cite{ferrari2013raman} At the level of an individual ribbon, this ratio may be used to tag the bonding motive of the carbon atoms surrounding the exciton. More precisely, the D-mode that is a second order process in graphene, becomes first-order in the case of GNRs, \cite{Fedotov2020Excitonic} explaining its high intensity in our STML spectra. The 677 cm$^{-1}$ peak, whose intensity varies differently than the rest of the spectra with tip position (Fig. \ref{fig2}A), is not reproduced by simple Franck-Condon simulations. Our closed-shell DFT calculations for ribbons of various lengths consistently reveal a normal mode at 692 cm$^{-1}$ that we tentatively associate with the experimental peak at 677 cm$^{-1}$. This mode is presenting a single antinode and is of odd symmetry (see the C-H$_{\rm IR}$ mode in Fig. \ref{fig4}B). We therefore suggest that this vibrational mode promotes non-adiabatic coupling (\textit{i.e.}, Herzberg-Teller active modes) between the localized exciton and higher-lying delocalized excitons, explaining the non-vanishing intensity of the 677 cm$^{-1}$ for the tip located on top of the middle of the GNR (Fig. \ref{fig2}A). Furthermore, in the experimental spectra one also distinguishes weak vibronic combs on the low energy side of the 677 cm$^{-1}$ and 1348 cm$^{-1}$ peaks that reproduce the shape and energy separation reported in Fig. \ref{fig3}. These peaks are therefore interpreted as combination bands replicating the low energy vibronic progression.\\
To conclude, our atomically resolved fluorescence measurements reveal sharp ($\approx$ 0.6 meV) emission from a long-lived dark exciton localized at the topological ends of ($7, m$)AGNRs. These localized emitting centers are coupled to one-dimensional acoustic phonon modes delocalized over the whole ribbon. Emitting centers localized in insulators and/or semiconductors, such as color centers or defects in solids,\cite{Gang2020} are often used as single- or entangled-photon sources of particular interest for quantum sensing and quantum technology applications. An advantage of the topologically localized centers in GNRs over more conventional solid-state quantum emitters is that one can tailor the number and the position of the photon sources through chemical engineering of the GNR short and long edges, providing thus an efficient path to tune inter-source coupling, and control the classical and quantum emission properties. An obvious next step of our work will be to identify the single-photon source character of the emission at the topologically localized centers and to characterize their performance. Besides, each topological end-state of ($7, m$)AGNRs hosts an unpaired electron and is therefore spin-polarized, providing organic nanoscale solutions for quantum schemes combining electronic, magnetic and photonic degrees of freedom. These GNRs can also be viewed as ideal atomically controlled platforms to identify, with atomic-scale spatial accuracy, the role of exciton-phonon coupling on the (de-)coherence of the quantum units. Eventually, one may consider to functionalize the two ends of the GNR with specifically chosen chromophores\cite{Li19} and determine how delocalized acoustic phonon modes affect the coherent coupling between the chromophore dipoles,\cite{zhang2016visualizing, Dolezal2022Real} somehow mimicking at the single molecular level the role of vibronic coupling in light harvesting complexes.\\
\section*{Acknowledgments}
We thank Anna Roslawska for fruitful discussions. We are grateful to Virginie Speisser and Michelangelo Rom\'eo for technical support. This project has received funding from the European Research Council (ERC) under the European Union's Horizon 2020 research and innovation program (grant agreement No 771850). The International Center for Frontier Research in Chemistry (FRC) is also acknowledged for financial support. This work is supported by ``Investissements d'Avenir'' LabEx PALM (ANR-10-LABX-0039-PALM).\\
\textbf{Supplementary Materials}\\
Materials and Methods\\
Supplementary text\\
Table S1\\
Fig S1 -- S4\\
\section{References}
\bibliography{Reference}

\begin{thebibliography}{53}
\expandafter\ifx\csname natexlab\endcsname\relax\def\natexlab#1{#1}\fi
\expandafter\ifx\csname bibnamefont\endcsname\relax
  \def\bibnamefont#1{#1}\fi
\expandafter\ifx\csname bibfnamefont\endcsname\relax
  \def\bibfnamefont#1{#1}\fi
\expandafter\ifx\csname citenamefont\endcsname\relax
  \def\citenamefont#1{#1}\fi
\expandafter\ifx\csname url\endcsname\relax
  \def\url#1{\texttt{#1}}\fi
\expandafter\ifx\csname urlprefix\endcsname\relax\def\urlprefix{URL }\fi
\providecommand{\bibinfo}[2]{#2}
\providecommand{\eprint}[2][]{\url{#2}}

\bibitem[{\citenamefont{Cai et~al.}(2010)\citenamefont{Cai, Ruffieux, Jaafar,
  Bieri, Braun, Blankenburg, Muoth, Seitsonen, Saleh, Feng
  et~al.}}]{cai2010atomically}
\bibinfo{author}{\bibfnamefont{J.}~\bibnamefont{Cai}},
  \bibinfo{author}{\bibfnamefont{P.}~\bibnamefont{Ruffieux}},
  \bibinfo{author}{\bibfnamefont{R.}~\bibnamefont{Jaafar}},
  \bibinfo{author}{\bibfnamefont{M.}~\bibnamefont{Bieri}},
  \bibinfo{author}{\bibfnamefont{T.}~\bibnamefont{Braun}},
  \bibinfo{author}{\bibfnamefont{S.}~\bibnamefont{Blankenburg}},
  \bibinfo{author}{\bibfnamefont{M.}~\bibnamefont{Muoth}},
  \bibinfo{author}{\bibfnamefont{A.~P.} \bibnamefont{Seitsonen}},
  \bibinfo{author}{\bibfnamefont{M.}~\bibnamefont{Saleh}},
  \bibinfo{author}{\bibfnamefont{X.}~\bibnamefont{Feng}}, \bibnamefont{et~al.},
  \bibinfo{journal}{Nature} \textbf{\bibinfo{volume}{466}},
  \bibinfo{pages}{470} (\bibinfo{year}{2010}).

\bibitem[{\citenamefont{Bo et~al.}(2021)\citenamefont{Bo, Zou, and
  Wang}}]{Bo2021Novel}
\bibinfo{author}{\bibfnamefont{W.}~\bibnamefont{Bo}},
  \bibinfo{author}{\bibfnamefont{Y.}~\bibnamefont{Zou}}, \bibnamefont{and}
  \bibinfo{author}{\bibfnamefont{J.}~\bibnamefont{Wang}}, \bibinfo{journal}{RSC
  Adv.} \textbf{\bibinfo{volume}{11}}, \bibinfo{pages}{33675}
  (\bibinfo{year}{2021}).

\bibitem[{\citenamefont{Houtsma et~al.}(2021)\citenamefont{Houtsma, de~la Rie,
  and St{\"o}hr}}]{Houtsma2021Atomically}
\bibinfo{author}{\bibfnamefont{R.~S.~K.} \bibnamefont{Houtsma}},
  \bibinfo{author}{\bibfnamefont{J.}~\bibnamefont{de~la Rie}},
  \bibnamefont{and}
  \bibinfo{author}{\bibfnamefont{M.}~\bibnamefont{St{\"o}hr}},
  \bibinfo{journal}{Chem. Soc. Rev.} \textbf{\bibinfo{volume}{50}},
  \bibinfo{pages}{6541} (\bibinfo{year}{2021}).

\bibitem[{\citenamefont{Miao et~al.}(2021)\citenamefont{Miao, Wang, Mu, and
  Wang}}]{Miao2021The}
\bibinfo{author}{\bibfnamefont{W.}~\bibnamefont{Miao}},
  \bibinfo{author}{\bibfnamefont{L.}~\bibnamefont{Wang}},
  \bibinfo{author}{\bibfnamefont{X.}~\bibnamefont{Mu}}, \bibnamefont{and}
  \bibinfo{author}{\bibfnamefont{J.}~\bibnamefont{Wang}}, \bibinfo{journal}{J.
  Mater. Chem. C} \textbf{\bibinfo{volume}{9}}, \bibinfo{pages}{13600}
  (\bibinfo{year}{2021}).

\bibitem[{\citenamefont{Wang et~al.}(2021)\citenamefont{Wang, Wang, Ma, Chen,
  Jiang, Chen, Xie, Li, and Wang}}]{wang2021graphene}
\bibinfo{author}{\bibfnamefont{H.}~\bibnamefont{Wang}},
  \bibinfo{author}{\bibfnamefont{H.~S.} \bibnamefont{Wang}},
  \bibinfo{author}{\bibfnamefont{C.}~\bibnamefont{Ma}},
  \bibinfo{author}{\bibfnamefont{L.}~\bibnamefont{Chen}},
  \bibinfo{author}{\bibfnamefont{C.}~\bibnamefont{Jiang}},
  \bibinfo{author}{\bibfnamefont{C.}~\bibnamefont{Chen}},
  \bibinfo{author}{\bibfnamefont{X.}~\bibnamefont{Xie}},
  \bibinfo{author}{\bibfnamefont{A.-P.} \bibnamefont{Li}}, \bibnamefont{and}
  \bibinfo{author}{\bibfnamefont{X.}~\bibnamefont{Wang}},
  \bibinfo{journal}{Nat. Rev. Phys.} \textbf{\bibinfo{volume}{3}},
  \bibinfo{pages}{791} (\bibinfo{year}{2021}).

\bibitem[{\citenamefont{Song et~al.}(2022)\citenamefont{Song, Ng, Edalatmanesh,
  Sol\'e, Peng, Koloren\v{c}, Sosnov\'a, Stetsovych, Su, Li et~al.}}]{Song2022}
\bibinfo{author}{\bibfnamefont{S.}~\bibnamefont{Song}},
  \bibinfo{author}{\bibfnamefont{P.~W.} \bibnamefont{Ng}},
  \bibinfo{author}{\bibfnamefont{S.}~\bibnamefont{Edalatmanesh}},
  \bibinfo{author}{\bibfnamefont{A.~P.} \bibnamefont{Sol\'e}},
  \bibinfo{author}{\bibfnamefont{X.}~\bibnamefont{Peng}},
  \bibinfo{author}{\bibfnamefont{J.}~\bibnamefont{Koloren\v{c}}},
  \bibinfo{author}{\bibfnamefont{Z.}~\bibnamefont{Sosnov\'a}},
  \bibinfo{author}{\bibfnamefont{O.}~\bibnamefont{Stetsovych}},
  \bibinfo{author}{\bibfnamefont{J.}~\bibnamefont{Su}},
  \bibinfo{author}{\bibfnamefont{J.}~\bibnamefont{Li}}, \bibnamefont{et~al.},
  \bibinfo{journal}{arXiv:2204.12880}  (\bibinfo{year}{2022}).

\bibitem[{\citenamefont{Koch et~al.}(2012)\citenamefont{Koch, Ample, Joachim,
  and Grill}}]{koch2012voltage}
\bibinfo{author}{\bibfnamefont{M.}~\bibnamefont{Koch}},
  \bibinfo{author}{\bibfnamefont{F.}~\bibnamefont{Ample}},
  \bibinfo{author}{\bibfnamefont{C.}~\bibnamefont{Joachim}}, \bibnamefont{and}
  \bibinfo{author}{\bibfnamefont{L.}~\bibnamefont{Grill}},
  \bibinfo{journal}{Nat. Nanotechnol.} \textbf{\bibinfo{volume}{7}},
  \bibinfo{pages}{713} (\bibinfo{year}{2012}).

\bibitem[{\citenamefont{Ruffieux et~al.}(2016)\citenamefont{Ruffieux, Wang,
  Yang, S{\'a}nchez-S{\'a}nchez, Liu, Dienel, Talirz, Shinde, Pignedoli,
  Passerone et~al.}}]{ruffieux2016surface}
\bibinfo{author}{\bibfnamefont{P.}~\bibnamefont{Ruffieux}},
  \bibinfo{author}{\bibfnamefont{S.}~\bibnamefont{Wang}},
  \bibinfo{author}{\bibfnamefont{B.}~\bibnamefont{Yang}},
  \bibinfo{author}{\bibfnamefont{C.}~\bibnamefont{S{\'a}nchez-S{\'a}nchez}},
  \bibinfo{author}{\bibfnamefont{J.}~\bibnamefont{Liu}},
  \bibinfo{author}{\bibfnamefont{T.}~\bibnamefont{Dienel}},
  \bibinfo{author}{\bibfnamefont{L.}~\bibnamefont{Talirz}},
  \bibinfo{author}{\bibfnamefont{P.}~\bibnamefont{Shinde}},
  \bibinfo{author}{\bibfnamefont{C.~A.} \bibnamefont{Pignedoli}},
  \bibinfo{author}{\bibfnamefont{D.}~\bibnamefont{Passerone}},
  \bibnamefont{et~al.}, \bibinfo{journal}{Nature}
  \textbf{\bibinfo{volume}{531}}, \bibinfo{pages}{489} (\bibinfo{year}{2016}).

\bibitem[{\citenamefont{Gr{\"o}ning et~al.}(2018)\citenamefont{Gr{\"o}ning,
  Wang, Yao, Pignedoli, Barin, Daniels, Cupo, Meunier, Feng, Narita
  et~al.}}]{groning2018engineering}
\bibinfo{author}{\bibfnamefont{O.}~\bibnamefont{Gr{\"o}ning}},
  \bibinfo{author}{\bibfnamefont{S.}~\bibnamefont{Wang}},
  \bibinfo{author}{\bibfnamefont{X.}~\bibnamefont{Yao}},
  \bibinfo{author}{\bibfnamefont{C.~A.} \bibnamefont{Pignedoli}},
  \bibinfo{author}{\bibfnamefont{G.~B.} \bibnamefont{Barin}},
  \bibinfo{author}{\bibfnamefont{C.}~\bibnamefont{Daniels}},
  \bibinfo{author}{\bibfnamefont{A.}~\bibnamefont{Cupo}},
  \bibinfo{author}{\bibfnamefont{V.}~\bibnamefont{Meunier}},
  \bibinfo{author}{\bibfnamefont{X.}~\bibnamefont{Feng}},
  \bibinfo{author}{\bibfnamefont{A.}~\bibnamefont{Narita}},
  \bibnamefont{et~al.}, \bibinfo{journal}{Nature}
  \textbf{\bibinfo{volume}{560}}, \bibinfo{pages}{209} (\bibinfo{year}{2018}).

\bibitem[{\citenamefont{Rizzo et~al.}(2018)\citenamefont{Rizzo, Veber, Cao,
  Bronner, Chen, Zhao, Rodriguez, Louie, Crommie, and
  Fischer}}]{rizzo2018topological}
\bibinfo{author}{\bibfnamefont{D.~J.} \bibnamefont{Rizzo}},
  \bibinfo{author}{\bibfnamefont{G.}~\bibnamefont{Veber}},
  \bibinfo{author}{\bibfnamefont{T.}~\bibnamefont{Cao}},
  \bibinfo{author}{\bibfnamefont{C.}~\bibnamefont{Bronner}},
  \bibinfo{author}{\bibfnamefont{T.}~\bibnamefont{Chen}},
  \bibinfo{author}{\bibfnamefont{F.}~\bibnamefont{Zhao}},
  \bibinfo{author}{\bibfnamefont{H.}~\bibnamefont{Rodriguez}},
  \bibinfo{author}{\bibfnamefont{S.~G.} \bibnamefont{Louie}},
  \bibinfo{author}{\bibfnamefont{M.~F.} \bibnamefont{Crommie}},
  \bibnamefont{and} \bibinfo{author}{\bibfnamefont{F.~R.}
  \bibnamefont{Fischer}}, \bibinfo{journal}{Nature}
  \textbf{\bibinfo{volume}{560}}, \bibinfo{pages}{204} (\bibinfo{year}{2018}).

\bibitem[{\citenamefont{Rizzo et~al.}(2020)\citenamefont{Rizzo, Veber, Jiang,
  McCurdy, Cao, Bronner, Chen, Louie, Fischer, and
  Crommie}}]{Daniel2020Inducing}
\bibinfo{author}{\bibfnamefont{D.~J.} \bibnamefont{Rizzo}},
  \bibinfo{author}{\bibfnamefont{G.}~\bibnamefont{Veber}},
  \bibinfo{author}{\bibfnamefont{J.}~\bibnamefont{Jiang}},
  \bibinfo{author}{\bibfnamefont{R.}~\bibnamefont{McCurdy}},
  \bibinfo{author}{\bibfnamefont{T.}~\bibnamefont{Cao}},
  \bibinfo{author}{\bibfnamefont{C.}~\bibnamefont{Bronner}},
  \bibinfo{author}{\bibfnamefont{T.}~\bibnamefont{Chen}},
  \bibinfo{author}{\bibfnamefont{S.~G.} \bibnamefont{Louie}},
  \bibinfo{author}{\bibfnamefont{F.~R.} \bibnamefont{Fischer}},
  \bibnamefont{and} \bibinfo{author}{\bibfnamefont{M.~F.}
  \bibnamefont{Crommie}}, \bibinfo{journal}{Science}
  \textbf{\bibinfo{volume}{369}}, \bibinfo{pages}{1597} (\bibinfo{year}{2020}).

\bibitem[{\citenamefont{Blackwell et~al.}(2021)\citenamefont{Blackwell, Zhao,
  Brooks, Zhu, Piskun, Wang, Delgado, Lee, Louie, and
  Fischer}}]{blackwell2021spin}
\bibinfo{author}{\bibfnamefont{R.~E.} \bibnamefont{Blackwell}},
  \bibinfo{author}{\bibfnamefont{F.}~\bibnamefont{Zhao}},
  \bibinfo{author}{\bibfnamefont{E.}~\bibnamefont{Brooks}},
  \bibinfo{author}{\bibfnamefont{J.}~\bibnamefont{Zhu}},
  \bibinfo{author}{\bibfnamefont{I.}~\bibnamefont{Piskun}},
  \bibinfo{author}{\bibfnamefont{S.}~\bibnamefont{Wang}},
  \bibinfo{author}{\bibfnamefont{A.}~\bibnamefont{Delgado}},
  \bibinfo{author}{\bibfnamefont{Y.-L.} \bibnamefont{Lee}},
  \bibinfo{author}{\bibfnamefont{S.~G.} \bibnamefont{Louie}}, \bibnamefont{and}
  \bibinfo{author}{\bibfnamefont{F.~R.} \bibnamefont{Fischer}},
  \bibinfo{journal}{Nature} \textbf{\bibinfo{volume}{600}},
  \bibinfo{pages}{647} (\bibinfo{year}{2021}).

\bibitem[{\citenamefont{Liu et~al.}(2022)\citenamefont{Liu, Fu, Liu, Narita,
  Samor\`i, Bonn, and Wang}}]{Liu2022Small}
\bibinfo{author}{\bibfnamefont{Z.}~\bibnamefont{Liu}},
  \bibinfo{author}{\bibfnamefont{S.}~\bibnamefont{Fu}},
  \bibinfo{author}{\bibfnamefont{X.}~\bibnamefont{Liu}},
  \bibinfo{author}{\bibfnamefont{A.}~\bibnamefont{Narita}},
  \bibinfo{author}{\bibfnamefont{P.}~\bibnamefont{Samor\`i}},
  \bibinfo{author}{\bibfnamefont{M.}~\bibnamefont{Bonn}}, \bibnamefont{and}
  \bibinfo{author}{\bibfnamefont{H.~I.} \bibnamefont{Wang}},
  \bibinfo{journal}{Adv. Sci.} \textbf{\bibinfo{volume}{n/a}},
  \bibinfo{pages}{2106055} (\bibinfo{year}{2022}).

\bibitem[{\citenamefont{Yang et~al.}(2007)\citenamefont{Yang, Cohen, and
  Louie}}]{Yang2007Excitonic}
\bibinfo{author}{\bibfnamefont{L.}~\bibnamefont{Yang}},
  \bibinfo{author}{\bibfnamefont{M.~L.} \bibnamefont{Cohen}}, \bibnamefont{and}
  \bibinfo{author}{\bibfnamefont{S.~G.} \bibnamefont{Louie}},
  \bibinfo{journal}{Nano Lett.} \textbf{\bibinfo{volume}{7}},
  \bibinfo{pages}{3112} (\bibinfo{year}{2007}).

\bibitem[{\citenamefont{Prezzi et~al.}(2008)\citenamefont{Prezzi, Varsano,
  Ruini, Marini, and Molinari}}]{Prezzi2008Optical}
\bibinfo{author}{\bibfnamefont{D.}~\bibnamefont{Prezzi}},
  \bibinfo{author}{\bibfnamefont{D.}~\bibnamefont{Varsano}},
  \bibinfo{author}{\bibfnamefont{A.}~\bibnamefont{Ruini}},
  \bibinfo{author}{\bibfnamefont{A.}~\bibnamefont{Marini}}, \bibnamefont{and}
  \bibinfo{author}{\bibfnamefont{E.}~\bibnamefont{Molinari}},
  \bibinfo{journal}{Phys. Rev. B} \textbf{\bibinfo{volume}{77}},
  \bibinfo{pages}{041404} (\bibinfo{year}{2008}).

\bibitem[{\citenamefont{Prezzi et~al.}(2011)\citenamefont{Prezzi, Varsano,
  Ruini, and Molinari}}]{Prezzi2011Quantum}
\bibinfo{author}{\bibfnamefont{D.}~\bibnamefont{Prezzi}},
  \bibinfo{author}{\bibfnamefont{D.}~\bibnamefont{Varsano}},
  \bibinfo{author}{\bibfnamefont{A.}~\bibnamefont{Ruini}}, \bibnamefont{and}
  \bibinfo{author}{\bibfnamefont{E.}~\bibnamefont{Molinari}},
  \bibinfo{journal}{Phys. Rev. B} \textbf{\bibinfo{volume}{84}},
  \bibinfo{pages}{041401} (\bibinfo{year}{2011}).

\bibitem[{\citenamefont{Alfonsi and Meneghetti}(2012)}]{Alfonsi2012Excitonic}
\bibinfo{author}{\bibfnamefont{J.}~\bibnamefont{Alfonsi}} \bibnamefont{and}
  \bibinfo{author}{\bibfnamefont{M.}~\bibnamefont{Meneghetti}},
  \bibinfo{journal}{New J. Phys.} \textbf{\bibinfo{volume}{14}},
  \bibinfo{pages}{053047} (\bibinfo{year}{2012}).

\bibitem[{\citenamefont{Attaccalite et~al.}(2017)\citenamefont{Attaccalite,
  Cannuccia, and Gr\"uning}}]{Attaccalite2017Excitonic}
\bibinfo{author}{\bibfnamefont{C.}~\bibnamefont{Attaccalite}},
  \bibinfo{author}{\bibfnamefont{E.}~\bibnamefont{Cannuccia}},
  \bibnamefont{and}
  \bibinfo{author}{\bibfnamefont{M.}~\bibnamefont{Gr\"uning}},
  \bibinfo{journal}{Phys. Rev. B} \textbf{\bibinfo{volume}{95}},
  \bibinfo{pages}{125403} (\bibinfo{year}{2017}).

\bibitem[{\citenamefont{Deilmann and Rohlfing}(2017)}]{Deilmann2017Huge}
\bibinfo{author}{\bibfnamefont{T.}~\bibnamefont{Deilmann}} \bibnamefont{and}
  \bibinfo{author}{\bibfnamefont{M.}~\bibnamefont{Rohlfing}},
  \bibinfo{journal}{Nano Lett.} \textbf{\bibinfo{volume}{17}},
  \bibinfo{pages}{6833} (\bibinfo{year}{2017}).

\bibitem[{\citenamefont{Cardoso et~al.}(2018)\citenamefont{Cardoso, Ferretti,
  and Prezzi}}]{cardoso2018termini}
\bibinfo{author}{\bibfnamefont{C.}~\bibnamefont{Cardoso}},
  \bibinfo{author}{\bibfnamefont{A.}~\bibnamefont{Ferretti}}, \bibnamefont{and}
  \bibinfo{author}{\bibfnamefont{D.}~\bibnamefont{Prezzi}},
  \bibinfo{journal}{Eur. Phys. J. B} \textbf{\bibinfo{volume}{91}},
  \bibinfo{pages}{286} (\bibinfo{year}{2018}).

\bibitem[{\citenamefont{Wu et~al.}(2020)\citenamefont{Wu, Wang, Liu, Zou, Shao,
  Shao, and Yam}}]{Wu2020Controlling}
\bibinfo{author}{\bibfnamefont{X.}~\bibnamefont{Wu}},
  \bibinfo{author}{\bibfnamefont{R.}~\bibnamefont{Wang}},
  \bibinfo{author}{\bibfnamefont{N.}~\bibnamefont{Liu}},
  \bibinfo{author}{\bibfnamefont{H.}~\bibnamefont{Zou}},
  \bibinfo{author}{\bibfnamefont{B.}~\bibnamefont{Shao}},
  \bibinfo{author}{\bibfnamefont{L.}~\bibnamefont{Shao}}, \bibnamefont{and}
  \bibinfo{author}{\bibfnamefont{C.}~\bibnamefont{Yam}},
  \bibinfo{journal}{Phys. Chem. Chem. Phys.} \textbf{\bibinfo{volume}{22}},
  \bibinfo{pages}{8277} (\bibinfo{year}{2020}).

\bibitem[{\citenamefont{Denk et~al.}(2014)\citenamefont{Denk, Hohage,
  Zeppenfeld, Cai, Pignedoli, S{\"o}de, Fasel, Feng, M{\"u}llen, Wang
  et~al.}}]{denk2014exciton}
\bibinfo{author}{\bibfnamefont{R.}~\bibnamefont{Denk}},
  \bibinfo{author}{\bibfnamefont{M.}~\bibnamefont{Hohage}},
  \bibinfo{author}{\bibfnamefont{P.}~\bibnamefont{Zeppenfeld}},
  \bibinfo{author}{\bibfnamefont{J.}~\bibnamefont{Cai}},
  \bibinfo{author}{\bibfnamefont{C.~A.} \bibnamefont{Pignedoli}},
  \bibinfo{author}{\bibfnamefont{H.}~\bibnamefont{S{\"o}de}},
  \bibinfo{author}{\bibfnamefont{R.}~\bibnamefont{Fasel}},
  \bibinfo{author}{\bibfnamefont{X.}~\bibnamefont{Feng}},
  \bibinfo{author}{\bibfnamefont{K.}~\bibnamefont{M{\"u}llen}},
  \bibinfo{author}{\bibfnamefont{S.}~\bibnamefont{Wang}}, \bibnamefont{et~al.},
  \bibinfo{journal}{Nat. Commun.} \textbf{\bibinfo{volume}{5}},
  \bibinfo{pages}{4253} (\bibinfo{year}{2014}).

\bibitem[{\citenamefont{Soavi et~al.}(2016)\citenamefont{Soavi, Dal~Conte,
  Manzoni, Viola, Narita, Hu, Feng, Hohenester, Molinari, Prezzi
  et~al.}}]{soavi2016exciton}
\bibinfo{author}{\bibfnamefont{G.}~\bibnamefont{Soavi}},
  \bibinfo{author}{\bibfnamefont{S.}~\bibnamefont{Dal~Conte}},
  \bibinfo{author}{\bibfnamefont{C.}~\bibnamefont{Manzoni}},
  \bibinfo{author}{\bibfnamefont{D.}~\bibnamefont{Viola}},
  \bibinfo{author}{\bibfnamefont{A.}~\bibnamefont{Narita}},
  \bibinfo{author}{\bibfnamefont{Y.}~\bibnamefont{Hu}},
  \bibinfo{author}{\bibfnamefont{X.}~\bibnamefont{Feng}},
  \bibinfo{author}{\bibfnamefont{U.}~\bibnamefont{Hohenester}},
  \bibinfo{author}{\bibfnamefont{E.}~\bibnamefont{Molinari}},
  \bibinfo{author}{\bibfnamefont{D.}~\bibnamefont{Prezzi}},
  \bibnamefont{et~al.}, \bibinfo{journal}{Nat. Commun.}
  \textbf{\bibinfo{volume}{7}}, \bibinfo{pages}{11010} (\bibinfo{year}{2016}).

\bibitem[{\citenamefont{Bronner et~al.}(2016)\citenamefont{Bronner, Gerbert,
  Broska, and Tegeder}}]{Bronner2016Excitonic}
\bibinfo{author}{\bibfnamefont{C.}~\bibnamefont{Bronner}},
  \bibinfo{author}{\bibfnamefont{D.}~\bibnamefont{Gerbert}},
  \bibinfo{author}{\bibfnamefont{A.}~\bibnamefont{Broska}}, \bibnamefont{and}
  \bibinfo{author}{\bibfnamefont{P.}~\bibnamefont{Tegeder}},
  \bibinfo{journal}{J. Phys. Chem. C} \textbf{\bibinfo{volume}{120}},
  \bibinfo{pages}{26168} (\bibinfo{year}{2016}).

\bibitem[{\citenamefont{Borin~Barin et~al.}(2019)\citenamefont{Borin~Barin,
  Fairbrother, Rotach, Bayle, Paillet, Liang, Meunier, Hauert, Dumslaff, Narita
  et~al.}}]{Barin2019Surface}
\bibinfo{author}{\bibfnamefont{G.}~\bibnamefont{Borin~Barin}},
  \bibinfo{author}{\bibfnamefont{A.}~\bibnamefont{Fairbrother}},
  \bibinfo{author}{\bibfnamefont{L.}~\bibnamefont{Rotach}},
  \bibinfo{author}{\bibfnamefont{M.}~\bibnamefont{Bayle}},
  \bibinfo{author}{\bibfnamefont{M.}~\bibnamefont{Paillet}},
  \bibinfo{author}{\bibfnamefont{L.}~\bibnamefont{Liang}},
  \bibinfo{author}{\bibfnamefont{V.}~\bibnamefont{Meunier}},
  \bibinfo{author}{\bibfnamefont{R.}~\bibnamefont{Hauert}},
  \bibinfo{author}{\bibfnamefont{T.}~\bibnamefont{Dumslaff}},
  \bibinfo{author}{\bibfnamefont{A.}~\bibnamefont{Narita}},
  \bibnamefont{et~al.}, \bibinfo{journal}{ACS Appl. Nano Mater.}
  \textbf{\bibinfo{volume}{2}}, \bibinfo{pages}{2184} (\bibinfo{year}{2019}).

\bibitem[{\citenamefont{Zhao et~al.}(2020)\citenamefont{Zhao, Barin, Cao,
  Overbeck, Darawish, Lyu, Drapcho, Wang, Dumslaff, Narita
  et~al.}}]{Zhao2020Optical}
\bibinfo{author}{\bibfnamefont{S.}~\bibnamefont{Zhao}},
  \bibinfo{author}{\bibfnamefont{G.~B.} \bibnamefont{Barin}},
  \bibinfo{author}{\bibfnamefont{T.}~\bibnamefont{Cao}},
  \bibinfo{author}{\bibfnamefont{J.}~\bibnamefont{Overbeck}},
  \bibinfo{author}{\bibfnamefont{R.}~\bibnamefont{Darawish}},
  \bibinfo{author}{\bibfnamefont{T.}~\bibnamefont{Lyu}},
  \bibinfo{author}{\bibfnamefont{S.}~\bibnamefont{Drapcho}},
  \bibinfo{author}{\bibfnamefont{S.}~\bibnamefont{Wang}},
  \bibinfo{author}{\bibfnamefont{T.}~\bibnamefont{Dumslaff}},
  \bibinfo{author}{\bibfnamefont{A.}~\bibnamefont{Narita}},
  \bibnamefont{et~al.}, \bibinfo{journal}{Nano Lett.}
  \textbf{\bibinfo{volume}{20}}, \bibinfo{pages}{1124} (\bibinfo{year}{2020}).

\bibitem[{\citenamefont{Tries et~al.}(2020)\citenamefont{Tries, Osella, Zhang,
  Xu, Ramanan, Kl{\"a}ui, Mai, Beljonne, and Wang}}]{Tries2020Experimental}
\bibinfo{author}{\bibfnamefont{A.}~\bibnamefont{Tries}},
  \bibinfo{author}{\bibfnamefont{S.}~\bibnamefont{Osella}},
  \bibinfo{author}{\bibfnamefont{P.}~\bibnamefont{Zhang}},
  \bibinfo{author}{\bibfnamefont{F.}~\bibnamefont{Xu}},
  \bibinfo{author}{\bibfnamefont{C.}~\bibnamefont{Ramanan}},
  \bibinfo{author}{\bibfnamefont{M.}~\bibnamefont{Kl{\"a}ui}},
  \bibinfo{author}{\bibfnamefont{Y.}~\bibnamefont{Mai}},
  \bibinfo{author}{\bibfnamefont{D.}~\bibnamefont{Beljonne}}, \bibnamefont{and}
  \bibinfo{author}{\bibfnamefont{H.~I.} \bibnamefont{Wang}},
  \bibinfo{journal}{Nano Lett.} \textbf{\bibinfo{volume}{20}},
  \bibinfo{pages}{2993} (\bibinfo{year}{2020}).

\bibitem[{\citenamefont{Senkovskiy et~al.}(2017)\citenamefont{Senkovskiy,
  Pfeiffer, Alavi, Bliesener, Zhu, Michel, Fedorov, German, Hertel, Haberer
  et~al.}}]{Senkovskiy2017Making}
\bibinfo{author}{\bibfnamefont{B.~V.} \bibnamefont{Senkovskiy}},
  \bibinfo{author}{\bibfnamefont{M.}~\bibnamefont{Pfeiffer}},
  \bibinfo{author}{\bibfnamefont{S.~K.} \bibnamefont{Alavi}},
  \bibinfo{author}{\bibfnamefont{A.}~\bibnamefont{Bliesener}},
  \bibinfo{author}{\bibfnamefont{J.}~\bibnamefont{Zhu}},
  \bibinfo{author}{\bibfnamefont{S.}~\bibnamefont{Michel}},
  \bibinfo{author}{\bibfnamefont{A.~V.} \bibnamefont{Fedorov}},
  \bibinfo{author}{\bibfnamefont{R.}~\bibnamefont{German}},
  \bibinfo{author}{\bibfnamefont{D.}~\bibnamefont{Hertel}},
  \bibinfo{author}{\bibfnamefont{D.}~\bibnamefont{Haberer}},
  \bibnamefont{et~al.}, \bibinfo{journal}{Nano Lett.}
  \textbf{\bibinfo{volume}{17}}, \bibinfo{pages}{4029} (\bibinfo{year}{2017}).

\bibitem[{\citenamefont{Pfeiffer et~al.}(2018)\citenamefont{Pfeiffer,
  Senkovskiy, Haberer, Fischer, Yang, Meerholz, Ando, Gr{\"u}neis, and
  Lindfors}}]{Pfeiffer2018Observation}
\bibinfo{author}{\bibfnamefont{M.}~\bibnamefont{Pfeiffer}},
  \bibinfo{author}{\bibfnamefont{B.~V.} \bibnamefont{Senkovskiy}},
  \bibinfo{author}{\bibfnamefont{D.}~\bibnamefont{Haberer}},
  \bibinfo{author}{\bibfnamefont{F.~R.} \bibnamefont{Fischer}},
  \bibinfo{author}{\bibfnamefont{F.}~\bibnamefont{Yang}},
  \bibinfo{author}{\bibfnamefont{K.}~\bibnamefont{Meerholz}},
  \bibinfo{author}{\bibfnamefont{Y.}~\bibnamefont{Ando}},
  \bibinfo{author}{\bibfnamefont{A.}~\bibnamefont{Gr{\"u}neis}},
  \bibnamefont{and} \bibinfo{author}{\bibfnamefont{K.}~\bibnamefont{Lindfors}},
  \bibinfo{journal}{Nano Lett.} \textbf{\bibinfo{volume}{18}},
  \bibinfo{pages}{7038} (\bibinfo{year}{2018}).

\bibitem[{\citenamefont{Ma et~al.}(2020)\citenamefont{Ma, Xiao, Puretzky, Wang,
  Mohsin, Huang, Liang, Luo, Lawrie, Gu et~al.}}]{Ma2020Engineering}
\bibinfo{author}{\bibfnamefont{C.}~\bibnamefont{Ma}},
  \bibinfo{author}{\bibfnamefont{Z.}~\bibnamefont{Xiao}},
  \bibinfo{author}{\bibfnamefont{A.~A.} \bibnamefont{Puretzky}},
  \bibinfo{author}{\bibfnamefont{H.}~\bibnamefont{Wang}},
  \bibinfo{author}{\bibfnamefont{A.}~\bibnamefont{Mohsin}},
  \bibinfo{author}{\bibfnamefont{J.}~\bibnamefont{Huang}},
  \bibinfo{author}{\bibfnamefont{L.}~\bibnamefont{Liang}},
  \bibinfo{author}{\bibfnamefont{Y.}~\bibnamefont{Luo}},
  \bibinfo{author}{\bibfnamefont{B.~J.} \bibnamefont{Lawrie}},
  \bibinfo{author}{\bibfnamefont{G.}~\bibnamefont{Gu}}, \bibnamefont{et~al.},
  \bibinfo{journal}{ACS Nano} \textbf{\bibinfo{volume}{14}},
  \bibinfo{pages}{5090} (\bibinfo{year}{2020}).

\bibitem[{\citenamefont{Chong et~al.}(2018)\citenamefont{Chong, Afshar-Imani,
  Scheurer, Cardoso, Ferretti, Prezzi, and Schull}}]{chong2018bright}
\bibinfo{author}{\bibfnamefont{M.~C.} \bibnamefont{Chong}},
  \bibinfo{author}{\bibfnamefont{N.}~\bibnamefont{Afshar-Imani}},
  \bibinfo{author}{\bibfnamefont{F.}~\bibnamefont{Scheurer}},
  \bibinfo{author}{\bibfnamefont{C.}~\bibnamefont{Cardoso}},
  \bibinfo{author}{\bibfnamefont{A.}~\bibnamefont{Ferretti}},
  \bibinfo{author}{\bibfnamefont{D.}~\bibnamefont{Prezzi}}, \bibnamefont{and}
  \bibinfo{author}{\bibfnamefont{G.}~\bibnamefont{Schull}},
  \bibinfo{journal}{Nano Lett.} \textbf{\bibinfo{volume}{18}},
  \bibinfo{pages}{175} (\bibinfo{year}{2018}).

\bibitem[{\citenamefont{Wang et~al.}(2016)\citenamefont{Wang, Talirz,
  Pignedoli, Feng, M{\"u}llen, Fasel, and Ruffieux}}]{wang2016giant}
\bibinfo{author}{\bibfnamefont{S.}~\bibnamefont{Wang}},
  \bibinfo{author}{\bibfnamefont{L.}~\bibnamefont{Talirz}},
  \bibinfo{author}{\bibfnamefont{C.~A.} \bibnamefont{Pignedoli}},
  \bibinfo{author}{\bibfnamefont{X.}~\bibnamefont{Feng}},
  \bibinfo{author}{\bibfnamefont{K.}~\bibnamefont{M{\"u}llen}},
  \bibinfo{author}{\bibfnamefont{R.}~\bibnamefont{Fasel}}, \bibnamefont{and}
  \bibinfo{author}{\bibfnamefont{P.}~\bibnamefont{Ruffieux}},
  \bibinfo{journal}{Nat. Commun.} \textbf{\bibinfo{volume}{7}},
  \bibinfo{pages}{11507} (\bibinfo{year}{2016}).

\bibitem[{\citenamefont{Qiu et~al.}(2003)\citenamefont{Qiu, Nazin, and
  Ho}}]{Qiu2003Vibrationally}
\bibinfo{author}{\bibfnamefont{X.~H.} \bibnamefont{Qiu}},
  \bibinfo{author}{\bibfnamefont{G.~V.} \bibnamefont{Nazin}}, \bibnamefont{and}
  \bibinfo{author}{\bibfnamefont{W.}~\bibnamefont{Ho}},
  \bibinfo{journal}{Science} \textbf{\bibinfo{volume}{299}},
  \bibinfo{pages}{542} (\bibinfo{year}{2003}).

\bibitem[{\citenamefont{Zhang et~al.}(2016)\citenamefont{Zhang, Luo, Zhang, Yu,
  Kuang, Zhang, Meng, Luo, Yang, Dong et~al.}}]{zhang2016visualizing}
\bibinfo{author}{\bibfnamefont{Y.}~\bibnamefont{Zhang}},
  \bibinfo{author}{\bibfnamefont{Y.}~\bibnamefont{Luo}},
  \bibinfo{author}{\bibfnamefont{Y.}~\bibnamefont{Zhang}},
  \bibinfo{author}{\bibfnamefont{Y.-J.} \bibnamefont{Yu}},
  \bibinfo{author}{\bibfnamefont{Y.-M.} \bibnamefont{Kuang}},
  \bibinfo{author}{\bibfnamefont{L.}~\bibnamefont{Zhang}},
  \bibinfo{author}{\bibfnamefont{Q.-S.} \bibnamefont{Meng}},
  \bibinfo{author}{\bibfnamefont{Y.}~\bibnamefont{Luo}},
  \bibinfo{author}{\bibfnamefont{J.-L.} \bibnamefont{Yang}},
  \bibinfo{author}{\bibfnamefont{Z.-C.} \bibnamefont{Dong}},
  \bibnamefont{et~al.}, \bibinfo{journal}{Nature}
  \textbf{\bibinfo{volume}{531}}, \bibinfo{pages}{623} (\bibinfo{year}{2016}).

\bibitem[{\citenamefont{Imada et~al.}(2016)\citenamefont{Imada, Miwa,
  Imai-Imada, Kawahara, Kimura, and Kim}}]{imada2016real}
\bibinfo{author}{\bibfnamefont{H.}~\bibnamefont{Imada}},
  \bibinfo{author}{\bibfnamefont{K.}~\bibnamefont{Miwa}},
  \bibinfo{author}{\bibfnamefont{M.}~\bibnamefont{Imai-Imada}},
  \bibinfo{author}{\bibfnamefont{S.}~\bibnamefont{Kawahara}},
  \bibinfo{author}{\bibfnamefont{K.}~\bibnamefont{Kimura}}, \bibnamefont{and}
  \bibinfo{author}{\bibfnamefont{Y.}~\bibnamefont{Kim}},
  \bibinfo{journal}{Nature} \textbf{\bibinfo{volume}{538}},
  \bibinfo{pages}{364} (\bibinfo{year}{2016}).

\bibitem[{\citenamefont{Doppagne et~al.}(2018)\citenamefont{Doppagne, Chong,
  Bulou, Boeglin, Scheurer, and Schull}}]{doppagne2018electrofluorochromism}
\bibinfo{author}{\bibfnamefont{B.}~\bibnamefont{Doppagne}},
  \bibinfo{author}{\bibfnamefont{M.~C.} \bibnamefont{Chong}},
  \bibinfo{author}{\bibfnamefont{H.}~\bibnamefont{Bulou}},
  \bibinfo{author}{\bibfnamefont{A.}~\bibnamefont{Boeglin}},
  \bibinfo{author}{\bibfnamefont{F.}~\bibnamefont{Scheurer}}, \bibnamefont{and}
  \bibinfo{author}{\bibfnamefont{G.}~\bibnamefont{Schull}},
  \bibinfo{journal}{Science} \textbf{\bibinfo{volume}{361}},
  \bibinfo{pages}{251} (\bibinfo{year}{2018}).

\bibitem[{\citenamefont{Dole\v{z}al et~al.}(2019)\citenamefont{Dole\v{z}al,
  Merino, Redondo, Ondi\v{c}, Cahl\'ik, and \v{S}vec}}]{dolezal19}
\bibinfo{author}{\bibfnamefont{J.}~\bibnamefont{Dole\v{z}al}},
  \bibinfo{author}{\bibfnamefont{P.}~\bibnamefont{Merino}},
  \bibinfo{author}{\bibfnamefont{J.}~\bibnamefont{Redondo}},
  \bibinfo{author}{\bibfnamefont{L.}~\bibnamefont{Ondi\v{c}}},
  \bibinfo{author}{\bibfnamefont{A.}~\bibnamefont{Cahl\'ik}}, \bibnamefont{and}
  \bibinfo{author}{\bibfnamefont{M.}~\bibnamefont{\v{S}vec}},
  \bibinfo{journal}{Nano Lett.} \textbf{\bibinfo{volume}{19}},
  \bibinfo{pages}{8605} (\bibinfo{year}{2019}).

\bibitem[{\citenamefont{Kimura et~al.}(2019)\citenamefont{Kimura, Miwa, Imada,
  Imai-Imada, Kawahara, Takeya, Kawai, Galperin, and
  Kim}}]{kimura2019selective}
\bibinfo{author}{\bibfnamefont{K.}~\bibnamefont{Kimura}},
  \bibinfo{author}{\bibfnamefont{K.}~\bibnamefont{Miwa}},
  \bibinfo{author}{\bibfnamefont{H.}~\bibnamefont{Imada}},
  \bibinfo{author}{\bibfnamefont{M.}~\bibnamefont{Imai-Imada}},
  \bibinfo{author}{\bibfnamefont{S.}~\bibnamefont{Kawahara}},
  \bibinfo{author}{\bibfnamefont{J.}~\bibnamefont{Takeya}},
  \bibinfo{author}{\bibfnamefont{M.}~\bibnamefont{Kawai}},
  \bibinfo{author}{\bibfnamefont{M.}~\bibnamefont{Galperin}}, \bibnamefont{and}
  \bibinfo{author}{\bibfnamefont{Y.}~\bibnamefont{Kim}},
  \bibinfo{journal}{Nature} \textbf{\bibinfo{volume}{570}},
  \bibinfo{pages}{210} (\bibinfo{year}{2019}).

\bibitem[{\citenamefont{Clair and de~Oteyza}(2019)}]{Clair2019Controlling}
\bibinfo{author}{\bibfnamefont{S.}~\bibnamefont{Clair}} \bibnamefont{and}
  \bibinfo{author}{\bibfnamefont{D.~G.} \bibnamefont{de~Oteyza}},
  \bibinfo{journal}{Chem. Rev.} \textbf{\bibinfo{volume}{119}},
  \bibinfo{pages}{4717} (\bibinfo{year}{2019}).

\bibitem[{\citenamefont{Doppagne et~al.}(2017)\citenamefont{Doppagne, Chong,
  Lorchat, Berciaud, Romeo, Bulou, Boeglin, Scheurer, and
  Schull}}]{Dop07Vibronic}
\bibinfo{author}{\bibfnamefont{B.}~\bibnamefont{Doppagne}},
  \bibinfo{author}{\bibfnamefont{M.~C.} \bibnamefont{Chong}},
  \bibinfo{author}{\bibfnamefont{E.}~\bibnamefont{Lorchat}},
  \bibinfo{author}{\bibfnamefont{S.}~\bibnamefont{Berciaud}},
  \bibinfo{author}{\bibfnamefont{M.}~\bibnamefont{Romeo}},
  \bibinfo{author}{\bibfnamefont{H.}~\bibnamefont{Bulou}},
  \bibinfo{author}{\bibfnamefont{A.}~\bibnamefont{Boeglin}},
  \bibinfo{author}{\bibfnamefont{F.}~\bibnamefont{Scheurer}}, \bibnamefont{and}
  \bibinfo{author}{\bibfnamefont{G.}~\bibnamefont{Schull}},
  \bibinfo{journal}{Phys. Rev. Lett.} \textbf{\bibinfo{volume}{118}},
  \bibinfo{pages}{127401} (\bibinfo{year}{2017}).

\bibitem[{\citenamefont{Kong et~al.}(2021)\citenamefont{Kong, Tian, Zhang, Yu,
  Jing, Zhang, Tian, Luo, Yang, Dong et~al.}}]{kong2021probing}
\bibinfo{author}{\bibfnamefont{F.-F.} \bibnamefont{Kong}},
  \bibinfo{author}{\bibfnamefont{X.-J.} \bibnamefont{Tian}},
  \bibinfo{author}{\bibfnamefont{Y.}~\bibnamefont{Zhang}},
  \bibinfo{author}{\bibfnamefont{Y.-J.} \bibnamefont{Yu}},
  \bibinfo{author}{\bibfnamefont{S.-H.} \bibnamefont{Jing}},
  \bibinfo{author}{\bibfnamefont{Y.}~\bibnamefont{Zhang}},
  \bibinfo{author}{\bibfnamefont{G.-J.} \bibnamefont{Tian}},
  \bibinfo{author}{\bibfnamefont{Y.}~\bibnamefont{Luo}},
  \bibinfo{author}{\bibfnamefont{J.-L.} \bibnamefont{Yang}},
  \bibinfo{author}{\bibfnamefont{Z.-C.} \bibnamefont{Dong}},
  \bibnamefont{et~al.}, \bibinfo{journal}{Nat. Commun.}
  \textbf{\bibinfo{volume}{12}}, \bibinfo{pages}{1280} (\bibinfo{year}{2021}).

\bibitem[{\citenamefont{van~der Lit et~al.}(2013)\citenamefont{van~der Lit,
  Boneschanscher, Vanmaekelbergh, Ij{\"a}s, Uppstu, Ervasti, Harju, Liljeroth,
  and Swart}}]{Lit13Suppression}
\bibinfo{author}{\bibfnamefont{J.}~\bibnamefont{van~der Lit}},
  \bibinfo{author}{\bibfnamefont{M.~P.} \bibnamefont{Boneschanscher}},
  \bibinfo{author}{\bibfnamefont{D.}~\bibnamefont{Vanmaekelbergh}},
  \bibinfo{author}{\bibfnamefont{M.}~\bibnamefont{Ij{\"a}s}},
  \bibinfo{author}{\bibfnamefont{A.}~\bibnamefont{Uppstu}},
  \bibinfo{author}{\bibfnamefont{M.}~\bibnamefont{Ervasti}},
  \bibinfo{author}{\bibfnamefont{A.}~\bibnamefont{Harju}},
  \bibinfo{author}{\bibfnamefont{P.}~\bibnamefont{Liljeroth}},
  \bibnamefont{and} \bibinfo{author}{\bibfnamefont{I.}~\bibnamefont{Swart}},
  \bibinfo{journal}{Nat. Commun.} \textbf{\bibinfo{volume}{4}},
  \bibinfo{pages}{2023} (\bibinfo{year}{2013}).

\bibitem[{\citenamefont{Talirz et~al.}(2013)\citenamefont{Talirz, S\"{o}de,
  Cai, Ruffieux, Blankenburg, Jafaar, Berger, Feng, M\"{u}llen, Passerone
  et~al.}}]{talirz2013termini}
\bibinfo{author}{\bibfnamefont{L.}~\bibnamefont{Talirz}},
  \bibinfo{author}{\bibfnamefont{H.}~\bibnamefont{S\"{o}de}},
  \bibinfo{author}{\bibfnamefont{J.}~\bibnamefont{Cai}},
  \bibinfo{author}{\bibfnamefont{P.}~\bibnamefont{Ruffieux}},
  \bibinfo{author}{\bibfnamefont{S.}~\bibnamefont{Blankenburg}},
  \bibinfo{author}{\bibfnamefont{R.}~\bibnamefont{Jafaar}},
  \bibinfo{author}{\bibfnamefont{R.}~\bibnamefont{Berger}},
  \bibinfo{author}{\bibfnamefont{X.}~\bibnamefont{Feng}},
  \bibinfo{author}{\bibfnamefont{K.}~\bibnamefont{M\"{u}llen}},
  \bibinfo{author}{\bibfnamefont{D.}~\bibnamefont{Passerone}},
  \bibnamefont{et~al.}, \bibinfo{journal}{J. Am. Chem. Soc.}
  \textbf{\bibinfo{volume}{135}}, \bibinfo{pages}{2060} (\bibinfo{year}{2013}).

\bibitem[{\citenamefont{Frisch et~al.}(2016)\citenamefont{Frisch, Trucks,
  Schlegel, Scuseria, Robb, Cheeseman, Scalmani, Barone, Petersson, Nakatsuji
  et~al.}}]{g16}
\bibinfo{author}{\bibfnamefont{M.~J.} \bibnamefont{Frisch}},
  \bibinfo{author}{\bibfnamefont{G.~W.} \bibnamefont{Trucks}},
  \bibinfo{author}{\bibfnamefont{H.~B.} \bibnamefont{Schlegel}},
  \bibinfo{author}{\bibfnamefont{G.~E.} \bibnamefont{Scuseria}},
  \bibinfo{author}{\bibfnamefont{M.~A.} \bibnamefont{Robb}},
  \bibinfo{author}{\bibfnamefont{J.~R.} \bibnamefont{Cheeseman}},
  \bibinfo{author}{\bibfnamefont{G.}~\bibnamefont{Scalmani}},
  \bibinfo{author}{\bibfnamefont{V.}~\bibnamefont{Barone}},
  \bibinfo{author}{\bibfnamefont{G.~A.} \bibnamefont{Petersson}},
  \bibinfo{author}{\bibfnamefont{H.}~\bibnamefont{Nakatsuji}},
  \bibnamefont{et~al.}, \emph{\bibinfo{title}{Gaussian~16 {R}evision {C}.01}}
  (\bibinfo{year}{2016}), \bibinfo{note}{{G}aussian Inc. Wallingford CT}.

\bibitem[{\citenamefont{Neuman et~al.}(2018)\citenamefont{Neuman, Esteban,
  Casanova, Garc\'{i}a-Vidal, and Aizpurua}}]{neuman2018nanolett}
\bibinfo{author}{\bibfnamefont{T.}~\bibnamefont{Neuman}},
  \bibinfo{author}{\bibfnamefont{R.}~\bibnamefont{Esteban}},
  \bibinfo{author}{\bibfnamefont{D.}~\bibnamefont{Casanova}},
  \bibinfo{author}{\bibfnamefont{F.~J.} \bibnamefont{Garc\'{i}a-Vidal}},
  \bibnamefont{and} \bibinfo{author}{\bibfnamefont{J.}~\bibnamefont{Aizpurua}},
  \bibinfo{journal}{Nano Letters} \textbf{\bibinfo{volume}{18}},
  \bibinfo{pages}{2358} (\bibinfo{year}{2018}).

\bibitem[{\citenamefont{Overbeck et~al.}(2019)\citenamefont{Overbeck, Barin,
  Daniels, Perrin, Braun, Sun, Darawish, De~Luca, Wang, Dumslaff
  et~al.}}]{overbeck19A}
\bibinfo{author}{\bibfnamefont{J.}~\bibnamefont{Overbeck}},
  \bibinfo{author}{\bibfnamefont{G.~B.} \bibnamefont{Barin}},
  \bibinfo{author}{\bibfnamefont{C.}~\bibnamefont{Daniels}},
  \bibinfo{author}{\bibfnamefont{M.~L.} \bibnamefont{Perrin}},
  \bibinfo{author}{\bibfnamefont{O.}~\bibnamefont{Braun}},
  \bibinfo{author}{\bibfnamefont{Q.}~\bibnamefont{Sun}},
  \bibinfo{author}{\bibfnamefont{R.}~\bibnamefont{Darawish}},
  \bibinfo{author}{\bibfnamefont{M.}~\bibnamefont{De~Luca}},
  \bibinfo{author}{\bibfnamefont{X.-Y.} \bibnamefont{Wang}},
  \bibinfo{author}{\bibfnamefont{T.}~\bibnamefont{Dumslaff}},
  \bibnamefont{et~al.}, \bibinfo{journal}{ACS Nano}
  \textbf{\bibinfo{volume}{13}}, \bibinfo{pages}{13083} (\bibinfo{year}{2019}).

\bibitem[{\citenamefont{Gillen et~al.}(2010)\citenamefont{Gillen, Mohr, and
  Maultzsch}}]{Gillen2010Raman}
\bibinfo{author}{\bibfnamefont{R.}~\bibnamefont{Gillen}},
  \bibinfo{author}{\bibfnamefont{M.}~\bibnamefont{Mohr}}, \bibnamefont{and}
  \bibinfo{author}{\bibfnamefont{J.}~\bibnamefont{Maultzsch}},
  \bibinfo{journal}{Phys. Status Solidi B} \textbf{\bibinfo{volume}{247}},
  \bibinfo{pages}{2941} (\bibinfo{year}{2010}).

\bibitem[{\citenamefont{Cong et~al.}(2019)\citenamefont{Cong, Li, Zhang, Lin,
  Wu, Liu, Venezuela, and Tan}}]{CONG201919Probing}
\bibinfo{author}{\bibfnamefont{X.}~\bibnamefont{Cong}},
  \bibinfo{author}{\bibfnamefont{Q.-Q.} \bibnamefont{Li}},
  \bibinfo{author}{\bibfnamefont{X.}~\bibnamefont{Zhang}},
  \bibinfo{author}{\bibfnamefont{M.-L.} \bibnamefont{Lin}},
  \bibinfo{author}{\bibfnamefont{J.-B.} \bibnamefont{Wu}},
  \bibinfo{author}{\bibfnamefont{X.-L.} \bibnamefont{Liu}},
  \bibinfo{author}{\bibfnamefont{P.}~\bibnamefont{Venezuela}},
  \bibnamefont{and} \bibinfo{author}{\bibfnamefont{P.-H.} \bibnamefont{Tan}},
  \bibinfo{journal}{Carbon} \textbf{\bibinfo{volume}{149}}, \bibinfo{pages}{19}
  (\bibinfo{year}{2019}).

\bibitem[{\citenamefont{Fedotov et~al.}(2020)\citenamefont{Fedotov, Rybkovskiy,
  Chernov, Obraztsova, and Obraztsova}}]{Fedotov2020Excitonic}
\bibinfo{author}{\bibfnamefont{P.~V.} \bibnamefont{Fedotov}},
  \bibinfo{author}{\bibfnamefont{D.~V.} \bibnamefont{Rybkovskiy}},
  \bibinfo{author}{\bibfnamefont{A.~I.} \bibnamefont{Chernov}},
  \bibinfo{author}{\bibfnamefont{E.~A.} \bibnamefont{Obraztsova}},
  \bibnamefont{and} \bibinfo{author}{\bibfnamefont{E.~D.}
  \bibnamefont{Obraztsova}}, \bibinfo{journal}{J. Phys. Chem. C}
  \textbf{\bibinfo{volume}{124}}, \bibinfo{pages}{25984}
  (\bibinfo{year}{2020}).

\bibitem[{\citenamefont{Ferrari and Basko}(2013)}]{ferrari2013raman}
\bibinfo{author}{\bibfnamefont{A.~C.} \bibnamefont{Ferrari}} \bibnamefont{and}
  \bibinfo{author}{\bibfnamefont{D.~M.} \bibnamefont{Basko}},
  \bibinfo{journal}{Nat. Nanotechnol.} \textbf{\bibinfo{volume}{8}},
  \bibinfo{pages}{235} (\bibinfo{year}{2013}).

\bibitem[{\citenamefont{Zhang et~al.}(2020)\citenamefont{Zhang, Cheng, Chou,
  and Gali}}]{Gang2020}
\bibinfo{author}{\bibfnamefont{G.}~\bibnamefont{Zhang}},
  \bibinfo{author}{\bibfnamefont{Y.}~\bibnamefont{Cheng}},
  \bibinfo{author}{\bibfnamefont{J.-P.} \bibnamefont{Chou}}, \bibnamefont{and}
  \bibinfo{author}{\bibfnamefont{A.}~\bibnamefont{Gali}},
  \bibinfo{journal}{Appl. Phys. Rev.} \textbf{\bibinfo{volume}{7}},
  \bibinfo{pages}{031308} (\bibinfo{year}{2020}).

\bibitem[{\citenamefont{Li et~al.}(2019)\citenamefont{Li, Friedrich, Merino,
  de~Oteyza, Pe\~na, Jacob, and Pascual}}]{Li19}
\bibinfo{author}{\bibfnamefont{J.}~\bibnamefont{Li}},
  \bibinfo{author}{\bibfnamefont{N.}~\bibnamefont{Friedrich}},
  \bibinfo{author}{\bibfnamefont{N.}~\bibnamefont{Merino}},
  \bibinfo{author}{\bibfnamefont{D.~G.} \bibnamefont{de~Oteyza}},
  \bibinfo{author}{\bibfnamefont{D.}~\bibnamefont{Pe\~na}},
  \bibinfo{author}{\bibfnamefont{D.}~\bibnamefont{Jacob}}, \bibnamefont{and}
  \bibinfo{author}{\bibfnamefont{J.~I.} \bibnamefont{Pascual}},
  \bibinfo{journal}{Nano Lett.} \textbf{\bibinfo{volume}{19}},
  \bibinfo{pages}{3288} (\bibinfo{year}{2019}).

\bibitem[{\citenamefont{Dole\v{z}al et~al.}(2022)\citenamefont{Dole\v{z}al,
  Canola, Hapala, de~Campos~Ferreira, Merino, and \v{S}vec}}]{Dolezal2022Real}
\bibinfo{author}{\bibfnamefont{J.}~\bibnamefont{Dole\v{z}al}},
  \bibinfo{author}{\bibfnamefont{S.}~\bibnamefont{Canola}},
  \bibinfo{author}{\bibfnamefont{P.}~\bibnamefont{Hapala}},
  \bibinfo{author}{\bibfnamefont{R.~C.} \bibnamefont{de~Campos~Ferreira}},
  \bibinfo{author}{\bibfnamefont{P.}~\bibnamefont{Merino}}, \bibnamefont{and}
  \bibinfo{author}{\bibfnamefont{M.}~\bibnamefont{\v{S}vec}},
  \bibinfo{journal}{ACS Nano} \textbf{\bibinfo{volume}{16}},
  \bibinfo{pages}{1082} (\bibinfo{year}{2022}).

\end{thebibliography}
\end{document}